\documentstyle[aps,prl,multicol,epsf]{revtex}

\begin{document}

\title{{\bf Scale Dependent Dimension in the Forest Fire Model}}
\author{Kan Chen and Per Bak}
\address{Department of Computational Science, National
University of Singapore, Singapore 117543 and\\
Niels Bohr Institute, Blegdamsvej 17, 2100 Copenhagen, Denmark}

\date{\today }
\maketitle

\begin{abstract}

The forest fire model is a reaction-diffusion model where energy,
in the form of trees, is injected uniformly, and burned
(dissipated) locally. We show that the spatial distribution of
fires forms a novel geometric structure where the fractal
dimension varies continuously with the length scale. In the three
dimensional model, the dimensions varies from zero to three,
proportional with $log(l)$, as the length scale increases from $l
\sim 1$ to a correlation length $l=\xi$. Beyond the correlation
length, which diverges with the growth rate $p$ as ${\xi} \propto
p^{-2/3}$, the distribution becomes homogeneous. We suggest that
this picture applies to the ``intermediate range'' of turbulence
where it provides a natural interpretation of the extended
scaling that has been observed at small length scales.
Unexpectedly, it might also be applicable to the spatial
distribution of luminous matter in the universe. In the
two-dimensional version, the dimension increases to $D=1$ at a
length scale $l \sim 1/p$, where there is a  cross-over to
homogeneity, i. e. a jump from $D=1$ to $D=2$.
\end{abstract}


\begin{multicols}{2}
\section{introduction}
Systems undergoing continuous phase transitions are usually
described in terms of a correlation length ${\xi}$ which diverges
as the critical point is approached. For length scales below the
correlation length, the geometrical structure is self-similar,
with a unique fractal dimension which is independent of $l$ for
$l\ll \xi$. For larger length scales the structure is uniform.
Self-similarity can be conveniently described as a fixed point of
a renormalization group transformation. This behaviour includes
self-organised critical phenomena, where the correlation length
diverges as a power-law when the driving rate vanishes.

We propose a new geometric form for non-equilibrium systems,
where the dimension of the dissipative field varies gradually
from $D = 0$  at  the smallest scale, dominated by point-like
objects, to $D = 3$, or bulk-like, at some finite  correlation
length ${\xi}$. The density remains uniform for length scales
exceeding ${\xi}$. Thus, as one steps further and further
backwards, and consider things at a larger and larger length
scale $l$, different classes of fractal objects are observed,
ranging from points to lines to walls, and finally to a
homogeneous set.

The new form of scaling appears in our study of the simple forest
fire model (FFM)\cite{BCT}, which was proposed in an attempt to
throw light on the nature of the spatial distribution of energy
dissipation in fully developed turbulence. It is a discrete model
defined on a lattice in the best traditions of Ising-like models
used to study equilibrium phase transitions. In turbulence,
energy is injected at a large scale, and dissipated at the
smallest scale. In our vision, the forest-fires would represent
the intermittency observed in turbulence, and power-law spatial
and temporal correlations would naturally occur if the model
operates at the self-organised critical state.

It turns out that, while the model indeed has a correlation
length that diverges as the growth rate $p$ goes to zero, there
is no fractal self-similarity below the correlation length, as is
usually the case for critical systems, self-organised or not. The
criticality can therefore not be described in terms of a fixed
point in the Wilson sense. For the three-dimensional version, we
find that the length scale dependent dimension $D(l)$ (for
$l<\xi$) is given by a remarkably simple equation,

\begin{equation}
\label{eq_1}
 D(l) \sim 3\log(l/l_0)/\log(\xi/l_0).
\end{equation}

\noindent where $l_0$ is a length of order unity. In two
dimensions, the exponent increases gradually to $D=1$ at the
correlation length, where there is a normal crossover to two
dimensional homogeneity. Whether or not one would actually call
this ``critical'' behaviour is a matter of taste.

While this novel geometric structure was not what we were looking
for, it may nevertheless turn out to represent the actual scaling
in turbulence. In the scaling regime, or inertial range, the
energy dissipation field in turbulence is known to be homogeneous
with great accuracy. However, there appears to be an
``intermediate range'' \cite{Frisch}, rather than a single lower
cut-off
 length, where the interesting intermittent behaviour that we usually
associate with turbulence takes place. The correlations do not
follow power-laws, but are characterised by smoothly varying
effective exponents. Only the relative moments follow
scale-independent ratios. This can perhaps be interpreted as a
scale dependent exponent for the dimension of the dissipative
field, as observed in our model.

On a quite different front, there has been much controversy about
the spatial distribution of luminous matter in the universe. The
apparent hierarchical structure has lead to many researchers to
believe that the distribution could be self-similar, or fractal
\cite{Pietronero}. However, the uniformity of the background
radiation requires that the universe is homogeneous at the
largest scale, contradicting the simple self-similar scaling
picture. Perhaps the luminous matter in the universe obeys this
new type of geometry, which unifies both observations. While we
hesitate to claim that the universe should be viewed as one giant
forest fire, we do suggest that the novel scaling picture may
represent a quite general geometrical form for non-equilibrium
dissipative systems.

\section{Numerical results}
The forest fire model is defined as follows. On a $d$-dimensional
lattice, trees are grown randomly at a rate $p$. During a time
unit, trees burn down (leaving room for new trees), and ignite
their neighbours. After a transient period, the system enters a
statistically stationary state with a complex distribution of
fires (and forests). This state is the object of our
investigation.

Despite the glaring simplicity of the FFM, reaching an
understanding of its properties of has turned out to be an
elusive goal. Grassberger and Kantz \cite{Grassberger1} argued
that there could be no criticality since the dynamics is simply
that of domain walls moving with a finite mean velocity, burning
everything in their wake. The motion of the walls  gives rise to
local oscillations in the fire intensity. This view was more or
less accepted by the community, and interest in the model
dropped. In the mean time, Drossel and Schwabl \cite{Drossel}
invented a different version of the FFM where fires are injected
at a small but finite rate, which exhibits non-controversial
conventional SOC \cite{BTW}. It can be exactly solved in one
dimension \cite{PB} \cite{Drossel2} in terms of a cascade process.

However, a few years ago, Johansen \cite{Johansen} found that the
time between two fronts is not limited by the growth rate, since
the fire-walls can propagate without burning much of the forest.
In fact, the fraction of trees that burn vanishes as $p
\rightarrow 0$. The process should be seen as a percolation-like
process rather than as solid fire fronts. Soon after, Broker and
Grassberger \cite{Grassberger2} demonstrated that the periodicity
of local oscillations in the fire density diverges with a
non-trivial exponent, as $p \rightarrow 0$.

Our picture of the $D=3$ FFM does not involve the concept of walls
or spirals: they only exist as well-defined quantities in the
minds of the observer when the system is viewed at a particular
scale. In two dimensions, walls are still meaningful,
representing the largest coherent structures.

We simulated both two and three dimensional systems with sizes up
to $2048^2$ and $1024^3$, respectively. Half a million time steps
were used to collect the statistics for each simulation. Special
care is needed for the simulation of a sustained forest fire due
to the fact that for a given system size $L$, the fire dies out
if the rate is not sufficiently high. This feature turns out to
be important for applications to the spread of diseases, and to
the existence of luminous matter in the universe. If artificially
restarted (say, by just adding a single fire), the system often
goes into a state with global oscillations. Typically we can only
study system sizes which are much larger than the correlation
length. This contrasts to conventional critical phenomena, where
information for length scales up to the size of the system can be
obtained even for systems which are smaller than the correlation
length. Finite size scaling allows one to derive critical
exponents from studying such systems. Here, in contrast, the
entire dynamics collapses as soon as the correlation length
reaches the system size. Thus, the range of scales $l$ that we
can study is squeezed by the inequalities $1 < l < \xi << L$,
making the calculation numerically demanding despite the
simplicity of the model. Our simulations involved more than a
billion sites!

The average amount of fires $n(l)$ within boxes of size $l$ which
contain fires was measured. Figure 1(a) shows $\log(n)$ vs
$\log(l)$ for a wide range of $p$ and $L=128, 256, 516, 1024$ for
$d=3$. There is no linear regime, indicating that there is no
well-defined fractal dimension, in contrast to our original
claim. The slopes of curves generally increase with $l$, and
saturate at a  value of $3$ at larger $l$ values, indicating that
the distribution becomes uniform beyond a length scale which we
identify as the correlation length. Also, $n$ increases with $p$
since the number of fires in average must be equal to the growth
rate in the stationary state.

A unified description can be obtained by plotting
$\log(n)/\log(l)$ vs $\log(l)/\log(\xi)$ for the same data
(Figure 1(b)), where good data collapse, involving length scales
extending over three orders of magnitude, is obtained when
choosing $\xi = (0.77p)^{-2/3}$ ($\nu=2/3$). Actually, a slightly
better fit is obtained if the lower cut-off is allowed to depend
on $p$, so that in principle there are two adjustable exponents.
Figure 1(c) shows $\log(n/n_0)/\log(l/l_0)$ vs
$\log(l/l_0)/\log(\xi/l_0)$ with $l_0 \propto p^{0.03}$.

The data collapse implies that $\log(n/n_0)/\log(l/l_0)$ can be
written in the form

\begin{equation}
 \log(n)/\log(l/l_0) = f[\log(l/l_0)/\log(\xi/l_0)],
\end{equation}

\noindent where the collapsed curve represents the function $f$.
It turns out to be instructive to define the derivative

\begin{equation}
D(l)=d\log(n)/d\log(l) = \alpha[\log(l/l_0)/\log(\xi/l_0)],
\end{equation}

\noindent which can be thought of as a length scale dependent
fractal dimension. This quantity as shown in Figure 2. The
function $\alpha$ is given by $\alpha(x)=f(x)+xf'(x)$. The curve
has an interesting and unusual shape: the function is linear for
length scales up to the correlation length where there is a sharp
kink. Beyond the correlation length, the value of $d$ is $3$: the
system is homogeneous for length scales beyond the correlation
length. The data collapse shown in Figure 1(b-c) and 2 indicates
bona fide scaling, in the sense that there is only one relevant
length $\xi/l_0)$ in the system, although of a quite novel and
unique nature, without self-similarity under a change of scale.

Thus, we arrive at the extremely simple Eq.~(\ref{eq_1}), which is
our main result. This shows how the apparent dimension increases
as the forest is viewed at larger and larger distances. At the
smallest scales, the fires are zero-dimensional and appear
point-like and isolated. As the scale increases, the dimension
increases logarithmically until at the correlation length it
becomes equal to three. It would be interesting to have a
computer-generated graphical visualisation of this change of
dimension.

The amount of fires within a box of size $l$ becomes

\begin{equation}
\log(n) \sim \Bigg({3\over 2}{\log(l/l_0)\over
\log(\xi/l_0)}\Bigg)\log(l/l_0) \quad .
\end{equation}

The exponent $\nu$ can be derived analytically by an argument of
energy conservation. The number of fires $n(\xi)$ in a box of
size $\xi$ times the number of boxes of that size, $(L/\xi)^d$,
scales as $pL^d$. Since the fractal dimension $D(l)$ is linearly
dependent on $\log(l)$, we have $n(\xi) = \xi^{d/2}$ (ignoring
the small $p$-dependence of $l_0$); this leads to $\nu = 2/d$.

It is interesting that the correlation function throughout the
scaling region $l << \xi$ is influenced by both the correlation
length and the smallest length scale of dissipation. For example,
one can estimate the correlation length by measuring the increase
of dimensionality from one small length scale to another. In
contrast, for conventional critical phenomena, the properties up
to the correlation length are those of the critical state, and for
$l << \xi$ there is no way to detect $\xi$. In addition, the
scaling form is invariant with respect to the transformation
$l\rightarrow l^\gamma$, $\xi \rightarrow \xi^{\gamma}$, and
$n(l) \rightarrow n(l)^{\gamma}$, which leaves the smallest
dissipation scale (at $l \sim 1$) unchanged.

Similar data for $d=2$ are shown in Figure 3. The dimension grows
from a small value close to zero at low length scales, and then
jumps from $D\sim1$ to $D=2$ at the correlation length. The
variation is non-linear, but nevertheless there is good data
collapses for $\nu = 1$. For a range of length scales less than
the correlation length, the dimension is close to 1, indicating
that walls form the largest coherent structures in the two
dimensional forest fire model. Again, at length scales greater
than the correlation length, the density becomes uniform, and
$D=2$. There is no range of length scales where the dimension is
between 1 and 2. We have also studied the higher moments of the
fire distribution, and similar scaling forms were found. The
picture of propagating fronts remain valid; the features seen at
lower length scales represent the internal structure of the
walls, which also scales with the correlation length.

The forest fire model was originally thought of as a toy model of
turbulence. Recently, deviations from fractal, or multi-fractal,
scaling has been interpreted as `extended self-similarity' in an
intermediate dissipative range \cite{Frisch}\cite{Gagne} between
the Kolmogorov length and the inertial range. Perhaps one might
understand this phenomenon geometrically in terms of the concept
of scale dependent dimension introduced here. In particular, data
presented by Benzi et al. \cite{Benzi} seem to indicate a
logarithmic dependence of scaling exponents versus length scale.
For homogeneous turbulence, the energy dissipation scales with
the third moment of the velocity differences. It would be
interesting to plot the deviations in the intermediate range in
order to check whether it could be accounted for by a scale
dependent dimension as given by Eq.~(\ref{eq_1}). Other
experiments showing a dimension depending on the Reynolds number
\cite{Sreeni}\cite{Goldburg} might alternatively be interpreted
as a scale dependent dimension. In any case, the view of
turbulence as a forest fire could constitute a powerful
metaphorical picture.

It is important to distinguish our geometric structure from that
of a random distribution of fires with the same density. Such a
structure would produce a $D(l)$ vs. $\log(l)$ curve which would
be zero until a length equal to the cube root of the density, i.
e. the average length. Then there would be a crossover jump to
$D=3$. Similarly, a fractal structure would show up as a constant
$D(l)$ up to the correlation length; then there would be a jump
to the Euclidean dimension.

{\it The structure of the universe:} Could it indeed be that the
universe operates at a similar self-organized state with the
spatial distribution of bright matter characterised by the
logarithmic scaling? We have analysed galaxy maps in order to
test this proposal, with very good agreement \cite{universe}. The
fit yields a correlation length of approximately $300 Mpc$, which
is outside the range of present galaxy catalogues, but will be
reached within a decade. It would be exciting to check if the
structure indeed is represented by the kink-curve, Figure 2.

So far, we have not reached an analytical understanding of this
simple, and completely novel, type of scaling. Traditional
renormalization group analysis based on rescaling of length scale
will not apply here, and it is our belief that a quite different
framework might be needed.

\section{Acknowledgements}
PB thanks the National University of Singapore  for great
hospitality. KC is grateful to the support and hospitality of
Niels Bohr Institute, where part of this work was done. We wish
to thank Maya Paczuski for helpful discussions and comments on
the manuscript.

\newpage

\noindent {\bf Figure captions}

\vspace{6mm}

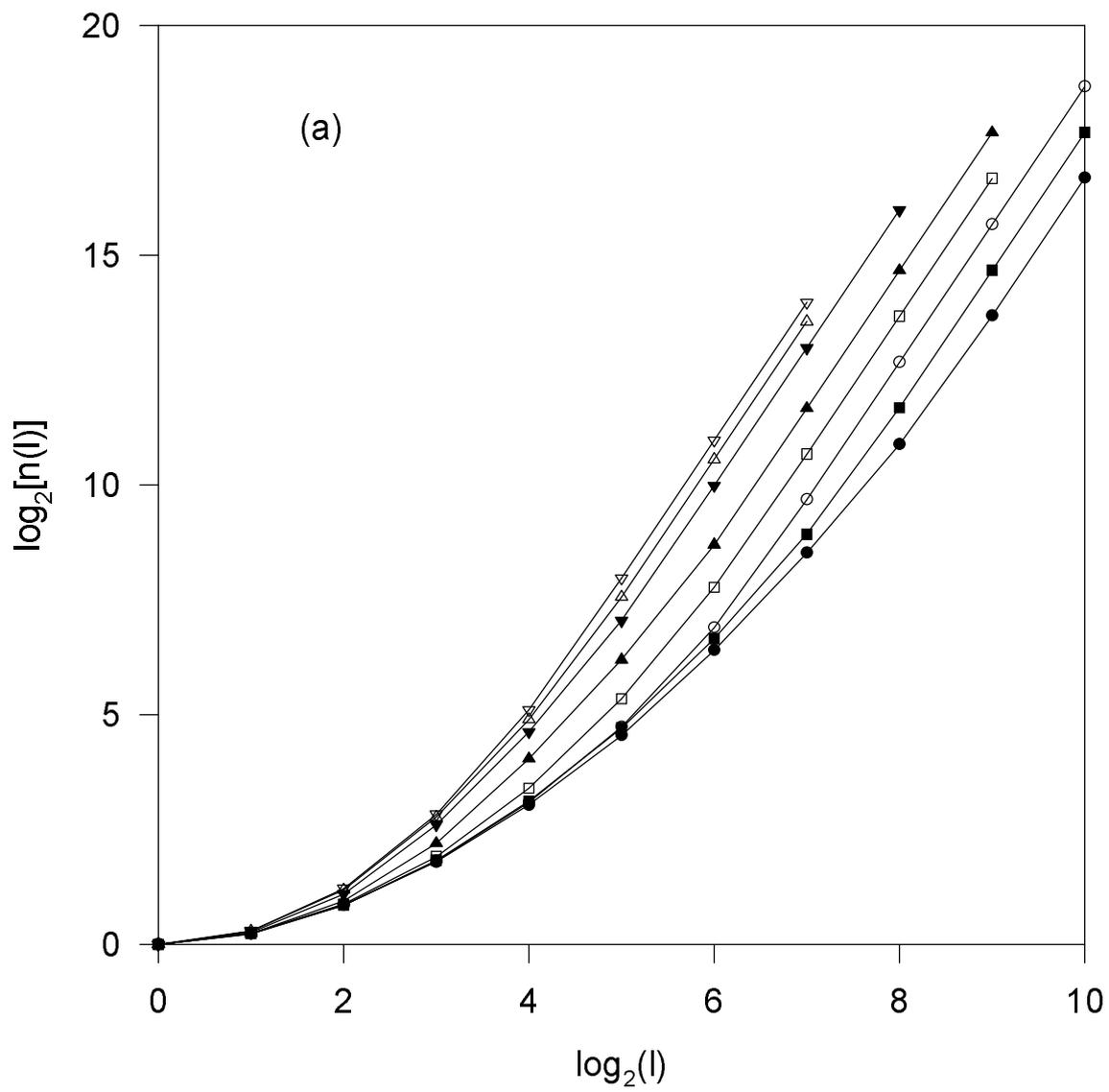
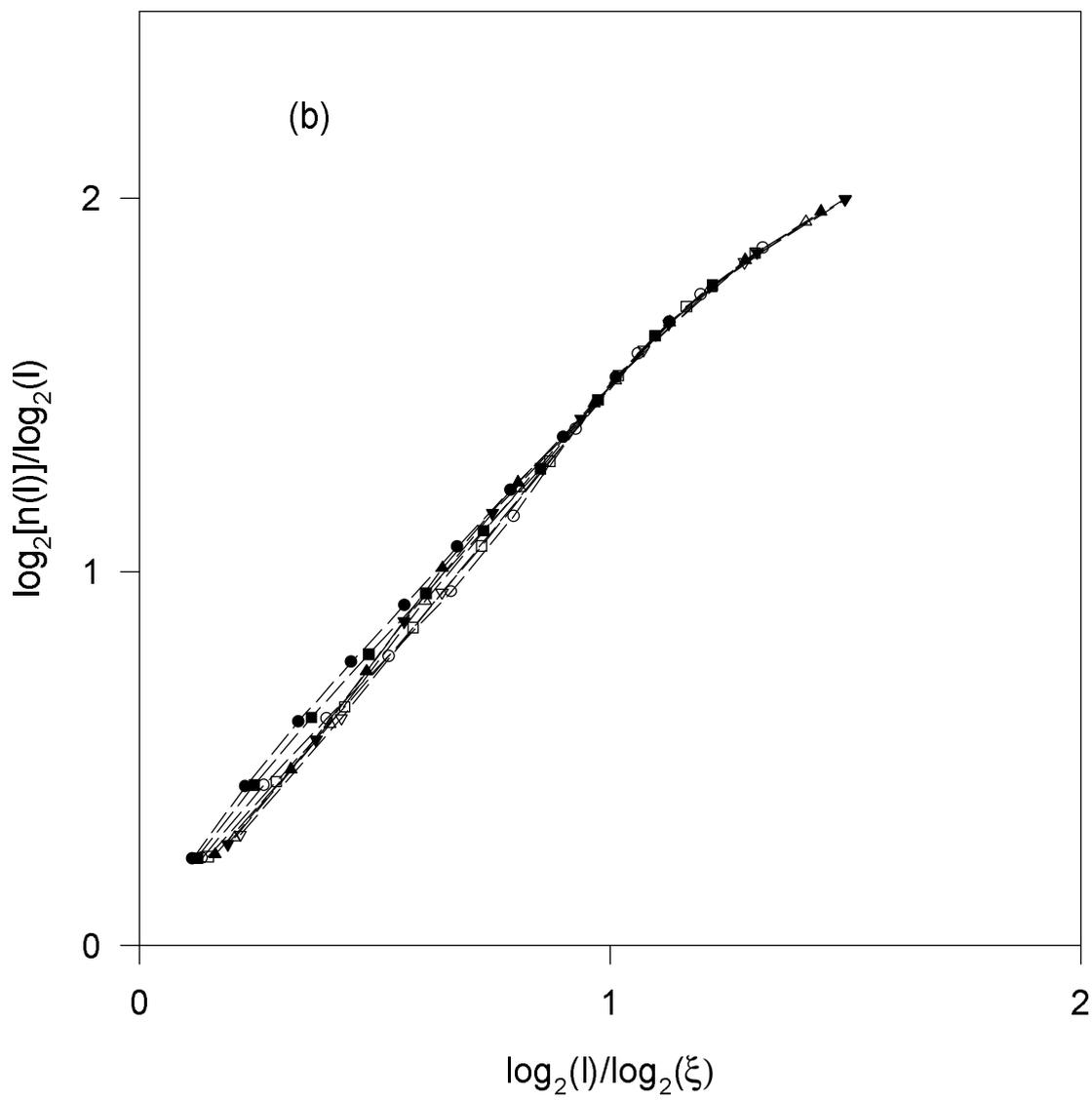
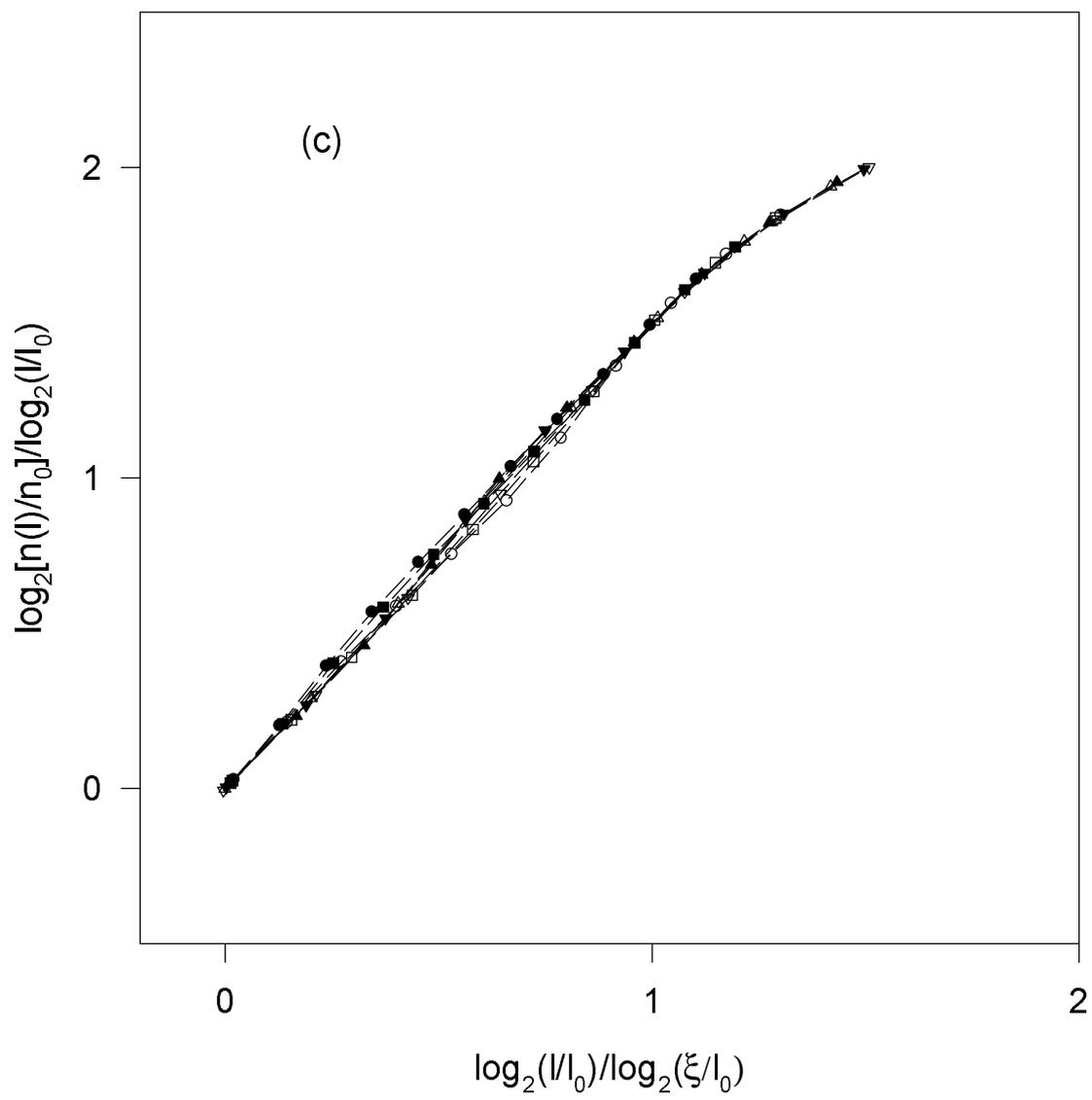
\noindent Figure 1. (a) $\log[n(l)]$ vs $\log(l)$ for various
system sizes and tree growth rates $p$ for the 3D forest fire
model: filled-circle ($1024^3, p=1.25\times 10^{-4}$),
filled-square ($1024^3, p=2.5\times 10^{-4}$), open-circle
($1024^3, p=5\times 10^{-4}$), open-square ($512^3, p=0.001$),
filled-triangle ($512^3, p=0.002$), inverted filled-triangle
($256^3, p=0.005$), open-triangle ($128^3,p=0.0075$), inverted
open-triangle ($128^3, p=0.01$). (b) $\log[n(l)]/\log(l)$ vs.
$\log(l)/\log(\xi)$ with the same set of data. The correlation
lengths are given by $\xi = (0.77p)^{-2/3}$. (c)
$\log[n(l)/n_0]/\log(l/l_0)$ vs $\log(l/l_0)/\log(\xi/l_0)$, with
the lower cut-off $l_0 \propto p^{0.03}$. \vspace{6mm}

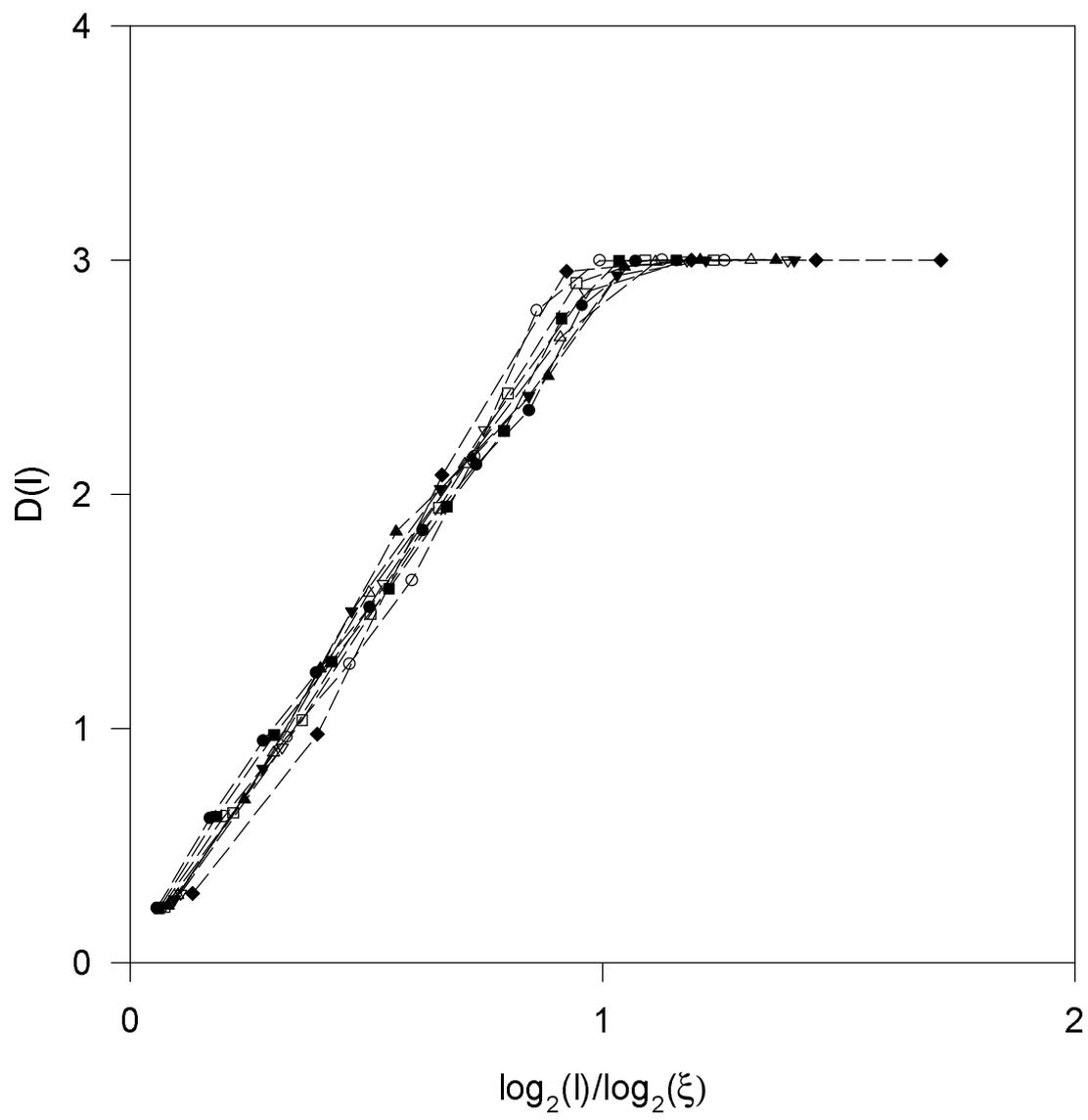
\noindent Figure 2. The length-dependent fractal dimension
$D(l)=d\log[n(l)]/d\log(l)$ (calculated from differences between
the data points in Figure 1) vs $\log(l)/\log(\xi)$ for the data
sets used in Figure 1, with $\xi = (0.77p)^{-2/3}$.

\vspace{6mm}

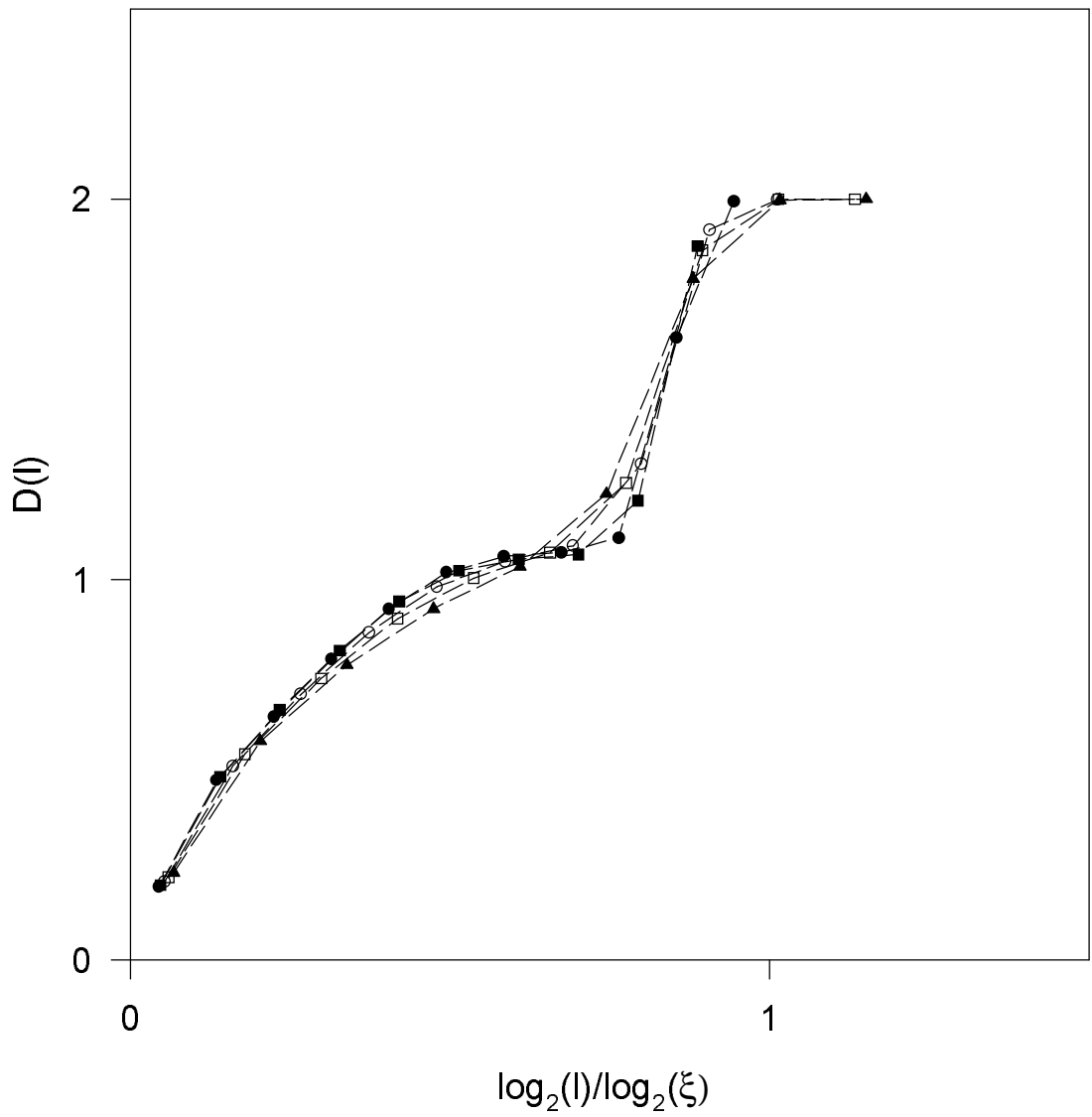
\noindent Figure 3. The length-dependent fractal dimension
$D(l)=d\log[n(l)]/d\log(l)$ vs $\log(l)/\log(\xi)$ for the 2d
forest fire model: filled-circle ($2048^2, p=0.00075$),
filled-square ($1024^2, p=0.001$), open-circle ($1024^2,
p=0.0025$), open-square ($1024^3, p=0.005$), filled-triangle
($516^2, p=0.01$). The correlation lengths used are given by
$\xi=(0.60p)^{-1}$.

\end{multicols}
\end{document}